\documentclass[preprint]{elsarticle} 

\usepackage[margin=2.7cm]{geometry}
\usepackage{amssymb}
\usepackage{amsmath}
\usepackage{lineno}
\usepackage{hyperref}

\begin{document}

\begin{frontmatter}
\title{New entropic inequalities for qubit and unimodal Gaussian states}

\author[n1,mi1]{J. A. L\'opez-Sald\'ivar\corref{cor1}}
\ead{julio.lopez@nucleares.unam.mx}
\author[n1]{ O. Casta\~nos}
\author[l1]{ M. A. Man'ko}
\author[l1,mi1]{ V. I. Man'ko}
 \address[n1]{Instituto de Ciencias Nucleares, Universidad Nacional
  Instituto de Ciencias Nucleares, Universidad Nacional Aut\'onoma de M\'exico, Apdo. Postal 70-543 M\'exico 04510 D.F.}
  \address[l1]{Lebedev Physical Institute, Leninskii Prospect 53,
Moscow 119991, Russia}
\address[mi1]{Moscow Institute of Physics and Technology (State University)
   Dolgoprudnyi, Moscow Region 141700, Russia
}
\cortext[cor1]{Corresponding author}

\begin{abstract}
The Tsallis relative
entropy $S_q (\hat{\rho},\hat{\sigma})$ measures the distance between two arbitrary
density matrices $\hat{\rho}$ and $\hat{\sigma}$. In this work the approximation to this quantity when $q=1+\delta$
($\delta\ll 1$) is obtained. It is shown that the resulting series is equal to the von Neumann relative
entropy when $\delta=0$. Analyzing the von Neumann relative entropy
for arbitrary $\hat{\rho}$ and a thermal equilibrium state
$\hat{\sigma}=e^{- \beta \hat{H}}/{\rm Tr}(e^{- \beta \hat{H}})$ is
possible to define a new inequality relating the energy, the
entropy, and the partition function of the system. From this
inequality, a parameter that measures the distance between the two
states is defined. This distance is calculated for a general qubit
system and for an arbitrary unimodal Gaussian state. In the qubit
case, the dependence on the purity of the system is studied for $T
\geq 0$ and also for $T<0$. In the Gaussian case, the general
partition function given a unimodal quadratic Hamiltonian is
calculated and the comparison of the thermal light state as a
thermal equilibrium state of the parametric amplifier is presented.
\end{abstract}


\end{frontmatter}

\section{Introduction}
The quantum aspects of thermodynamics have been subjects of intense
studies in recent review articles; see, e.g.,~\cite{huber,parrondo}.
There are many topics relating quantum information theory and
thermodynamics since the Maxwell's demon and the Landauer
principle~\cite{szilard,maruyama} to entanglement in many-body
systems~\cite{amico}. In particular, how concepts and techniques
from quantum mechanics or quantum information theory have been
applied to thermodynamics~\cite{esposito} and inversely how concepts
from thermodynamics are applied in the context of quantum
information~\cite{brandao,skrzypczyk,faist,yunger}.

The importance of the study of classical and quantum entropies as the ones defined by von Neumann~\cite{neumann} and Tsallis~\cite{tsallis1} for
quantum information systems has been increasing in the last decade.
In particular, new entropic inequalities are used to characterize
quantum correlations of qudit systems~\cite{nouvo}. They have lead
to nonlinear relations between unitary matrices which have been
checked in experiments of superconducting
qudits~\cite{fedorov,kiktenko,glushkov}.

The relative entropy~\cite{nielsen} ($\textrm{Tr}\,(\hat{\rho}\ln
\hat{\rho}-\hat{\rho}\ln\hat{\sigma})\geq 0 $), has been used to
describe finite systems approaching thermal
equilibrium~\cite{gaveau}; in those systems this quantity can be
interpreted as information needed to change an arbitrary density
matrix $\hat{\rho}$ to the thermal equilibrium
$\hat{\sigma}$~\cite{esposito}.

In the previous work~\cite{figueroa}, it was found that the sum of
the dimensionless energy $E(\hat{\rho},\hat{H})={\rm Tr}(\hat{\rho}
\hat{H})$ and entropy $S(\hat{\rho})=-{\rm Tr}(\hat{\rho} \ln
\hat{\rho})$ of the system is bounded by $\ln Z(\beta=-1, \hat{H})$,
where the partition function $Z(\beta,\hat{H})={\rm Tr}(e^{-\beta
\hat{H}})$  at $\beta=-1$ determines the value of the bound. As it
was shown in~\cite{nouvo}, the sum of the energy and the entropy
equals to the bound only if the density operator $\hat{\rho}$ of the
system equals to the thermal equilibrium state at temperature value
$T=-1$.

The aim of this work is to study the distance between and arbitrary
density matrix and a state approaching thermal equilibrium. In order
to do this, a new quantum inequality relating the free energy
($F(\hat{\rho},\hat{H}, T)=-T \ln (Z(\beta, \hat{H}))$) of an
arbitrary system with the corresponding partition function is
obtained. We apply the inequality to a spin-1/2 state and also for
an unimodal Gaussian state. We show that the bound in this
inequality is determined by the thermal equilibrium state~(TES) of
the system and thus a parameter for the distance can be defined.

\section{Tsallis relative entropy}
The Tsallis relative entropy introduced in \cite{hasegawa} and expressed as
\begin{equation}
S_q (\hat{\rho},\hat{\sigma})=\frac{1}{1-q}(1- {\rm Tr}(\hat{\rho}^q
\hat{\sigma}^{1-q}))
\label{tsallis}
\end{equation}
measures the distance between the density matrices $\hat{\rho}$ and
$\hat{\sigma}$. This quantity is zero iff the two are equal and
greater than zero in other case. When $q\rightarrow 1$ this
expression corresponds to the von Neumann relative entropy. If the
parameter $q$ approach $1$ ($q=1+\delta$, $\delta >0$, $\delta\ll
1$), Eq.~(\ref{tsallis}) can be written as a Taylor series of
$\delta$ as follows:
\begin{eqnarray}
S_{1+\delta} (\hat{\rho},\hat{\sigma})&=&{\rm Tr}(\hat{\rho}(\ln
\hat{\rho}-\ln \hat{\sigma}))+\frac{1}{2!}{\rm Tr}(\hat{\rho}(\ln
\hat{\rho}-\ln \hat{\sigma})^2) \delta+ \nonumber \\
&+& \frac{1}{3!}({\rm Tr}(\hat{\rho}(\ln \hat{\rho}-\ln \hat{\sigma})^3
+{\rm Tr}(\hat{\rho}(\ln\hat{\rho}\ln \hat{\sigma}-\ln \hat{\sigma}
\ln\hat{\rho})\ln \hat{\sigma})))\delta^2+ \mathcal{O}(\delta^3) \, ,
\end{eqnarray}
this expression can be checked using the expansion $\hat{\rho}^q=e^{q \ln \hat{\rho}}={\bf 1}+q \ln \hat{\rho}+q^2 \ln^2 \hat{\rho}/2!+\cdots$ and an analogous expression for $\hat{\sigma}$.
Using the relative entropy between two states ($\hat{\rho}$ and
$\hat{\sigma}$), where $\hat{\sigma}$ is a thermal equilibrium
state, it is possible to show the relation between the energy, the
entropy and the partition function. If the state $\hat{\sigma}$
corresponds to a mixed state dependent on a parameter $T$ (which it
may or may not be related with the temperature of the system), in
the form of
\begin{eqnarray}
\hat{\sigma} = \frac{e^{-\beta \hat{H}}}{\textrm{Tr}(e^{-\beta \hat{H}})} \, ,
\end{eqnarray}
then the positivity of the relative entropy defines an inequality
relation between the entropy, energy and the normalization factor of
$\hat{\sigma}$: Tr$(e^{-\beta \hat{H}})$. This relation for
$T>0$ has the following form:
\[
E(\hat{\rho},\hat{H})-T S(\hat{\rho}) \geq -T \ln (Z(\beta,\hat{H})) \, .
\]
This inequality must hold for any temperature $T$ for a fixed Hamiltonian.
Defining the parameter $\Delta$ as
\begin{equation}
\Delta = E(\hat{\rho},\hat{H})-T S(\hat{\rho})+ T \ln (Z(\beta,\hat{H}))  \geq 0\, ,
\label{ineq}
\end{equation}
then the difference between any state and a TES can be evaluated.

In this work, the study of the $\Delta$ parameter is presented for
an arbitrary qubit system and a general unimodal Gaussian state.

\section{Qubit system}
In a previous work~\cite{figueroa}, the entropy--energy relation for
extremal density matrices describing qudit states was analyzed. In
this section, a generalization of this relation is obtained for any
qubit system.

The most general density matrix for a qubit system is parametrized
by the Bloch vector $\mathbf{p}=(p_x, p_y, p_z)$, where each
component is given by the mean value of the Pauli matrices in a
system described by the density matrix $\hat{\rho}$, i.e., $p_x={\rm
Tr}(\hat{\rho}\, \hat{\sigma}_x)$, $p_y={\rm
Tr}(\hat{\rho}\,\hat{\sigma}_y)$, and $p_z={\rm
Tr}(\hat{\rho}\,\hat{\sigma}_z)$. In this case, the density matrix
$\hat{\rho}$ is given by the expression
\begin{equation}
\hat{\rho}=\frac{1}{2}\left( \begin{array}{cc}
1+p_z & p_x - i p_y \\
p_x+ip_y & 1-p_z
\end{array} \right) \, .
\end{equation}
In order to guaranty the positivity of the density matrix, the norm
of its Bloch vector must be bounded $0\leq \vert \mathbf{p} \vert
\leq 1 $; on the other hand, this norm also fixes the purity of the
state $\vert \mathbf{p} \vert = \sqrt{2 \, {\rm
Tr}(\hat{\rho}^2)-1}$, if $\vert \mathbf{p} \vert = 1$ the state is
pure and in other case the state is mixed. Also if $\vert \mathbf{p}
\vert=0$, the density matrix corresponds to the most mixed state.

In an analogous way, the Hamiltonian can be expressed in terms of
its own Bloch vector $\mathbf{h}=(h_1,h_2,h_3)$ as
\begin{equation}
\hat{H}=\frac{1}{2}\left( \begin{array}{cc}
h_0+h_3 & h_1 - i h_2 \\
h_1 + i h_2 & h_0- h_3
\end{array} \right) \, ,
\end{equation}
where $h_0={\rm Tr}(\hat{H})$.

\begin{figure}
\centering
(a)
\includegraphics[scale=0.25]{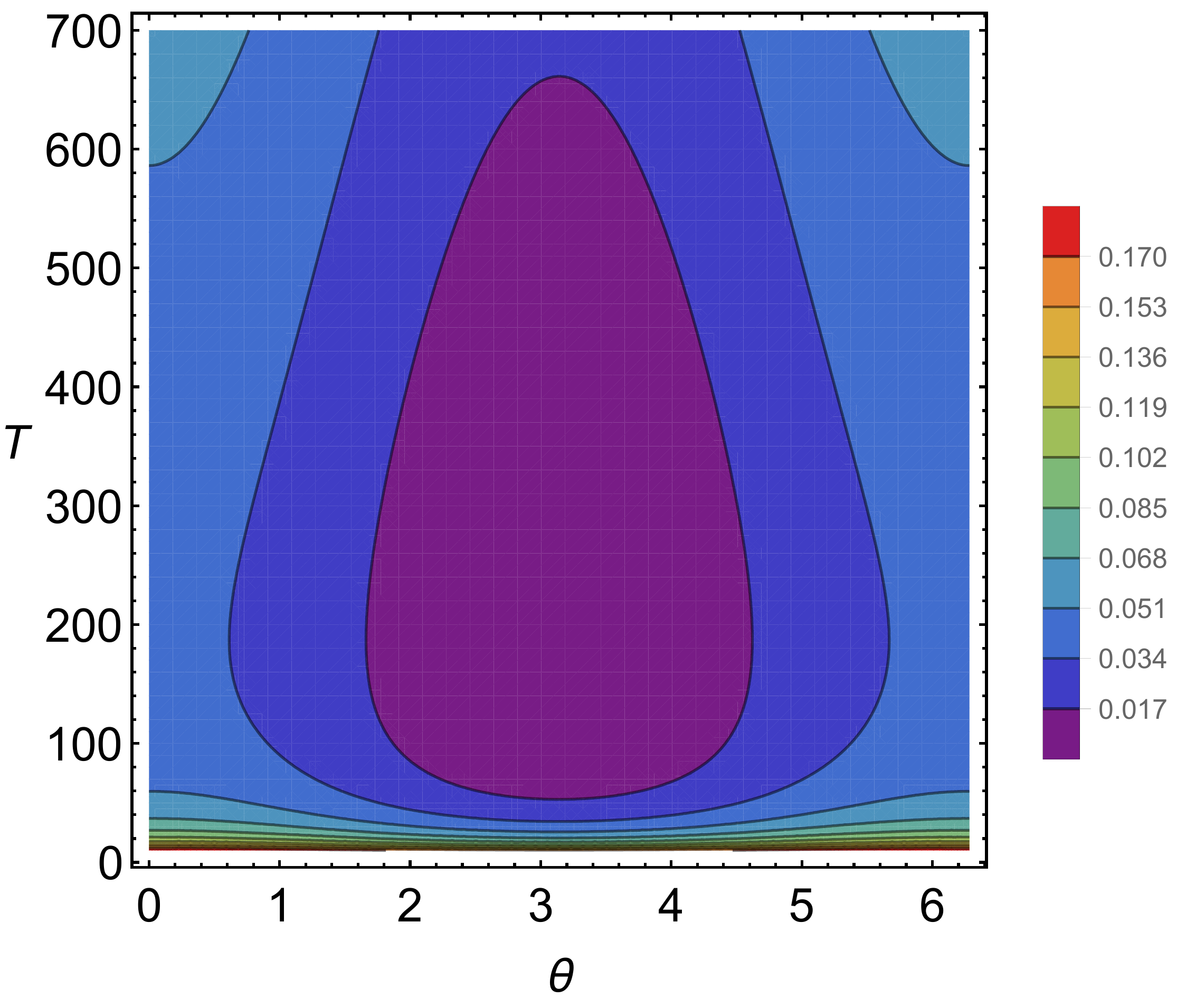}
(b)
\includegraphics[scale=0.25]{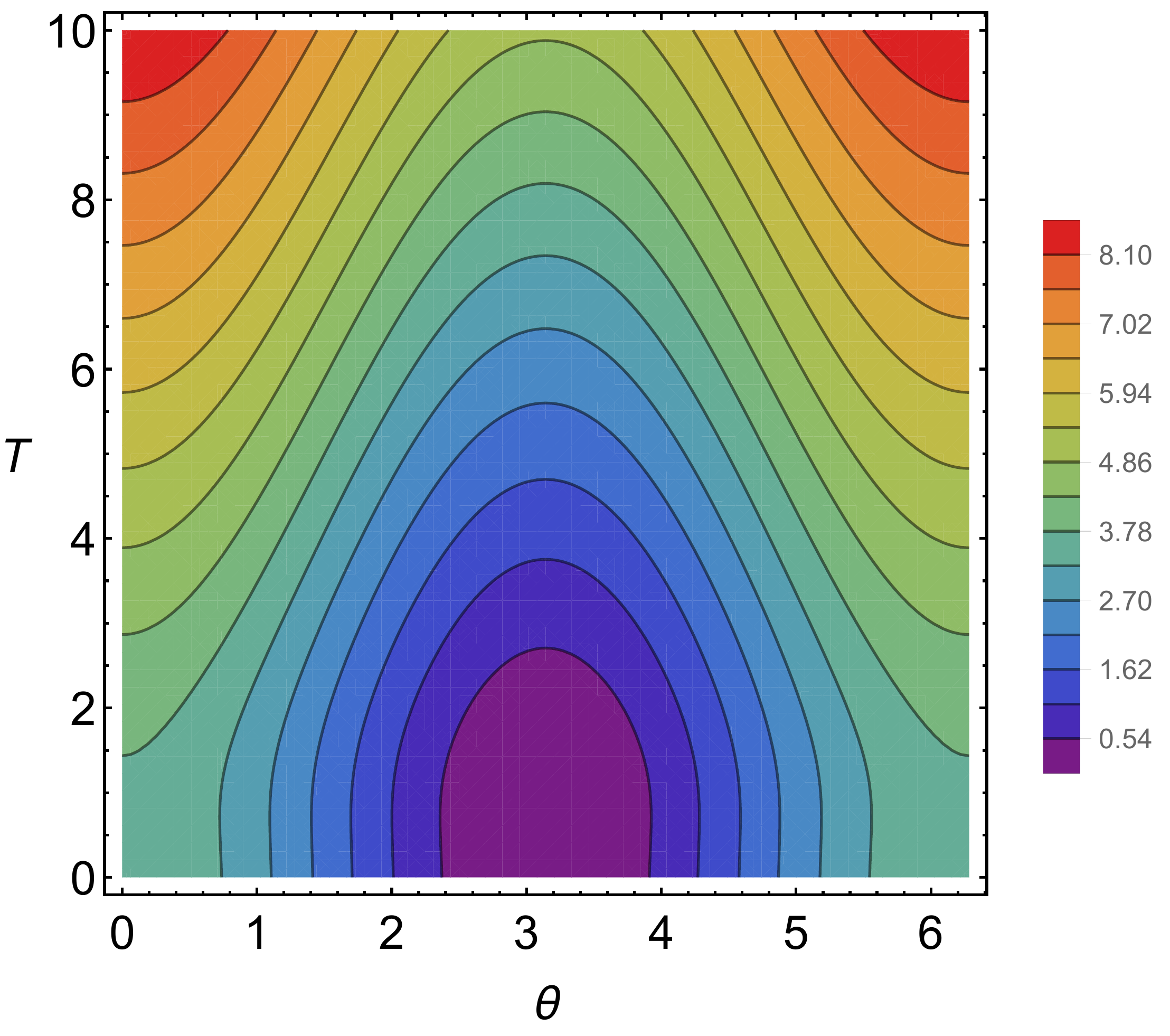}
\caption{(a) Contour plots of the $\Delta$ parameter as a function of
the $\theta$ angle for $T>0$ in a qubit system with $\vert \mathbf{p}
\vert = 0.01$ and $\vert \mathbf{h} \vert =\sqrt{14}$.
(b) $\vert \mathbf{p} \vert = 0.99$ .}
\label{fig:spinm}
\end{figure}

In the most general case, the mean value of the energy, the entropy
and the partition function of the system are given by the following
\begin{eqnarray}
& \langle \hat{H} \rangle = \frac{1}{2}(h_0+\mathbf{p}\cdot \mathbf{h}),
\quad S=-\frac{1}{2}\{(1-\vert \mathbf{p} \vert)\ln (1-\vert \mathbf{p}
\vert)+(1+\vert \mathbf{p} \vert)\ln (1+\vert \mathbf{p} \vert) \}, \nonumber \\
& Z(\beta, \hat{H})= e^{-\beta h_0 /2} \cosh (\beta \vert \mathbf{h} \vert /2) \, .
\end{eqnarray}
One can see that $\hat{\rho}$ corresponds to a thermal equilibrium
state ($\Delta=0$), when its Bloch vector and the Bloch vector of
the Hamiltonian follow the relation
\begin{equation}
\mathbf{p}=-\frac{\mathbf{h}}{\vert \mathbf{h}\vert} \tanh \left(
\frac{\beta \vert \mathbf{h} \vert}{2}\right) \, ,
\end{equation}
this is when the Bloch vectors are antiparallel and $T=\vert \mathbf{h} \vert/(2
\, { \rm arctanh} \vert \mathbf{p}\vert)$.  In fig.~(\ref{fig:spinm}a),
one can see the dependence of the $\Delta$ parameter as a function
of $\theta$ and $T$ for given $\vert \mathbf{p} \vert = 0.01$ and
$\vert \mathbf{h} \vert=\sqrt{14}$; in this case, $\Delta=0$ for
$T\approx 187$. A small $\vert \mathbf{p} \vert$ indicates a state
near the most mixed state (${\rm Tr}(\hat{\rho}^2)=1/2$). In
fig.~(\ref{fig:spinm}b), the plots for $\vert \mathbf{p} \vert=0.99$
(almost a pure state) can be seen, in that case $\Delta=0$ for
$T\approx 0.71$.

\begin{figure}
\centering
(a)
\includegraphics[scale=0.25]{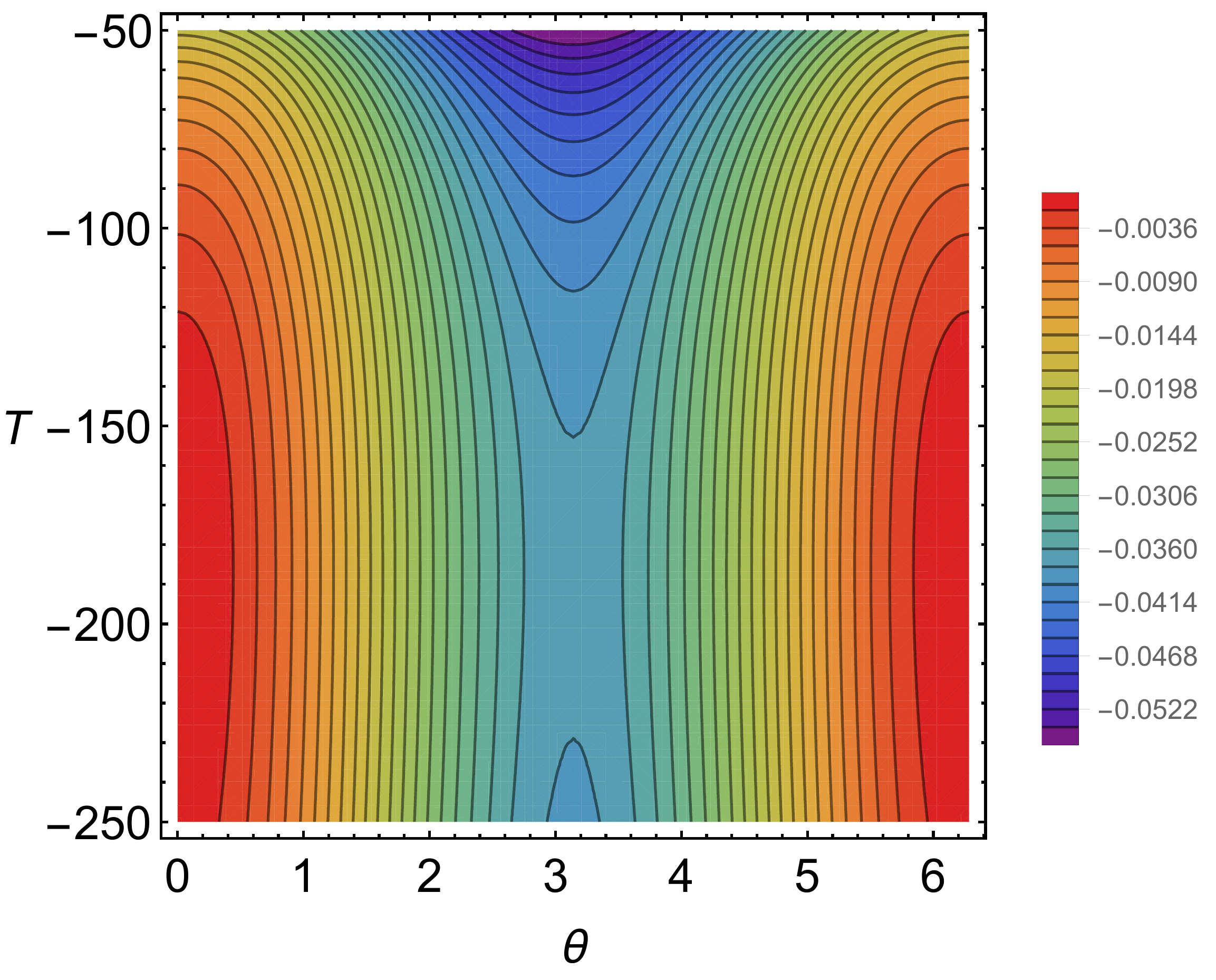}
(b)
\includegraphics[scale=0.25]{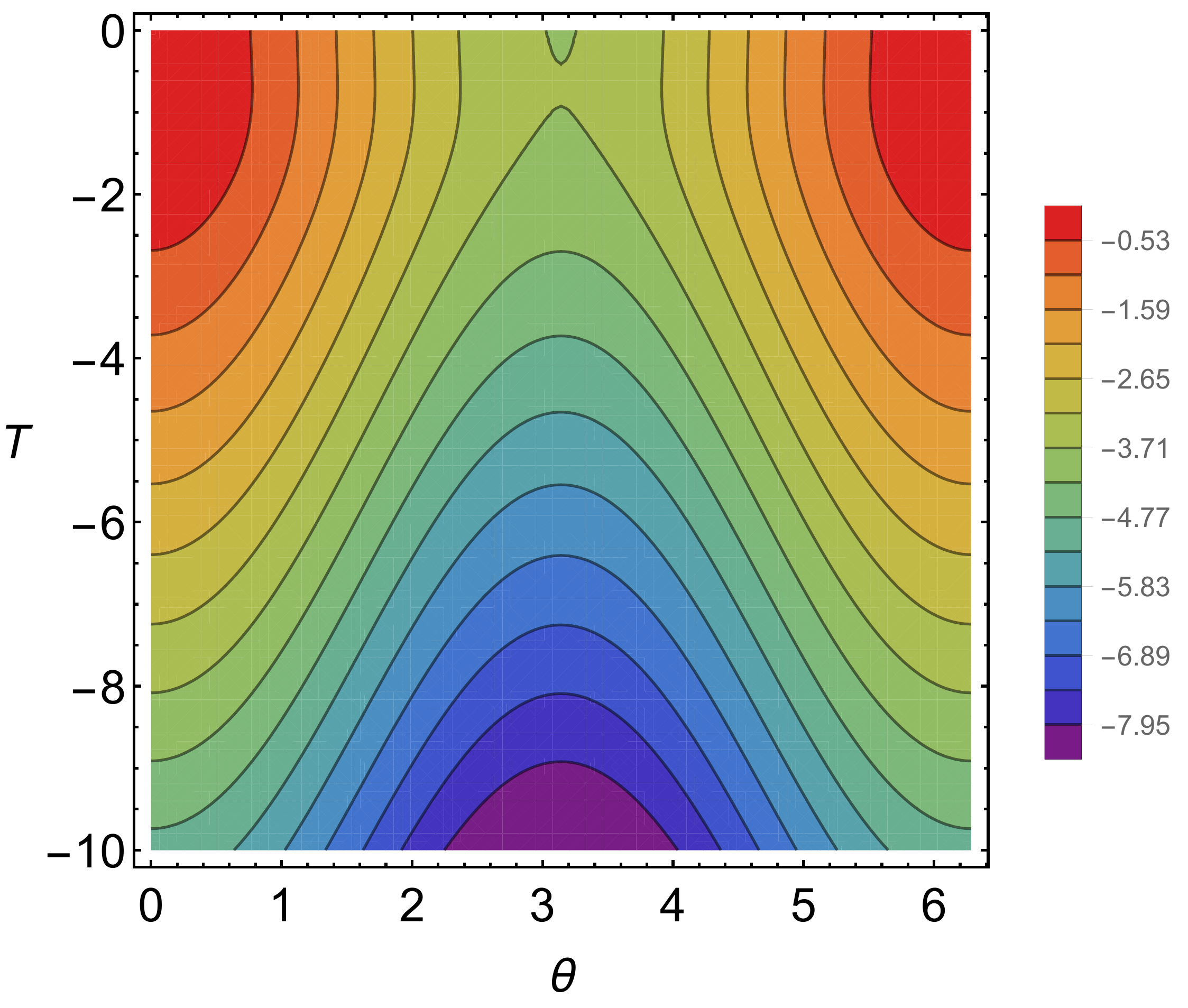}
\caption{(a) Contour plots of the $\Delta$ parameter as a function of
the $\theta$ angle for $T<0$ in a qubit system with $\vert \mathbf{p}
\vert = 0.01$ and $\vert \mathbf{h} \vert =\sqrt{14}$.
(b) $\vert \mathbf{p} \vert = 0.99$ .}
\label{fig:spinmt}
\end{figure}

It is possible to do the same analysis for $T < 0$; in that case,
the parameter $\Delta$ of~Eq.($\ref{ineq}$) must be negative. The
state $\hat{\rho}$ is a thermal equilibrium state when $\mathbf{p}$
and $\mathbf{h}$ are parallel for $T=-\vert \mathbf{h} \vert/(2 \, {
\rm arctanh} \vert \mathbf{p}\vert)$. The corresponding plots of
this parameter are presented in fig.~(\ref{fig:spinmt}a) for an
almost most mixed state ($\vert \mathbf{p} \vert=0.01$) and in
fig.~(\ref{fig:spinmt}b) for an almost pure state ($\vert \mathbf{p}
\vert=0.99$). In the case of the mixed state, $\Delta=0$ for
$T\approx -187$, for $\vert \mathbf{p} \vert=0.99$ (almost pure
state)  $\Delta=0$ for $T \approx -0.71$.

\section{Single-mode Gaussian states as thermal equilibrium states}
In recent years, the Gaussian states have become very important due
to their use in quantum information theory~\cite{weedbrook,adesso}.
In this section, we study the parameter $\Delta$ for a general
unimodal Gaussian state. The most general unimodal Gaussian
state~(GS) is described by the density matrix
\begin{eqnarray}
\hat{\rho} (q', q)= \mathcal{N} e^{-a_1 q^2-a_1^* q'^2+a_2 q q'+
 b_1 q + b_1^* q'}, \nonumber \\
 \mathcal{N}= \sqrt{ \frac{a_1+ a_1^*-a_2}{\pi}} \exp \left(
 \frac{-(b_1+ b_1^*)^2}{4(a_1+a_1^*-a_2)}\right) \, ,
 \label{gs}
\end{eqnarray}
where the normalization condition (${\rm Tr}(\hat{\rho})=1$) implies
that parameter $a_2 \in \mathbb{R}$ and $2 \textrm{Re} (a_1) > a_2$.
The coherent state, the squeezed vacuum state and the thermal light state
are examples of Gaussian states. The properties of these states are
determined by the covariance matrix of the system
$\boldsymbol{\sigma}$ and the mean values in the quadrature
components ($\hat{p}$, $\hat{q}$). For the general case of Eq.
(\ref{gs}), the covariance matrix is
\begin{equation}
\boldsymbol{\sigma}=\frac{1}{2(a_1+a_1^*-a_2)} \left( \begin{array}{cc}
4 \vert a_1 \vert^2 - a_2^2 & i(a_1^* -a_1) \\ i (a_1^* -a_1) &  1
\end{array} \right),
\end{equation}
while the mean values are
\begin{equation}
\langle \hat{p} \rangle = \frac{ a_2 \, {\rm Im(b_1)}-2 \,
{\rm Im}(a_1^* b_1)}{2(a_1+a_1^*-a_2)} , \ \ \langle \hat{q} \rangle
= \frac{b_1+b_1^*}{2(a_1+a_1^*-a_2)}\, .
\end{equation}

\subsection{Quadratic Hamiltonian}
It can be seen that a state of the form $e^{-\beta \hat{H}}/{\rm
Tr}(e^{-\beta \hat{H}})$ with $\hat{H}$ being quadratic is a
Gaussian state. Due to this, in order to describe the unimodal
Gaussian states as quasi-thermal equilibrium states, we obtain some
properties of quadratic Hamiltonians.

A general one-mode quadratic Hamiltonian can be written as
\begin{equation}
\hat{H} = \boldsymbol{\zeta} \boldsymbol{\Omega}
\widetilde{\boldsymbol{\zeta}} \, ,
\label{hamil}
\end{equation}
where $\boldsymbol{\zeta}=  (\hat{p}, \hat{q})$, $\boldsymbol{\Omega}=
\left( \begin{array}{cc} \omega_1 & \omega_2 \\ \omega_2^* & \omega_3
\end{array} \right)$, and the tilde means the transposition operation.
These parameters may be functions of the time.

It can be demonstrated that, for any Gaussian state with covariance
matrix $\boldsymbol{\sigma}$ and mean values $\langle
\boldsymbol{\zeta}\rangle = (\langle \hat{p}\rangle, \langle \hat{q}
\rangle)$, the mean value of the energy can be expressed as
\begin{equation}
E(\hat{\rho},\hat{H})=\langle \hat{H} \rangle = \textrm{Tr}
( \boldsymbol{\Omega}  \boldsymbol{\sigma})+\langle \boldsymbol{\zeta}
\rangle  \boldsymbol{\Omega}  \langle \widetilde{\boldsymbol{\zeta}}
\rangle + \textrm{Im}(\omega_{2}) \, .
\label{mean}
\end{equation}
where $\textrm{Im}(\omega_{2})$ denotes the imaginary part of $\omega_2$.

The partition function of the system $Z(\beta, \hat{H})= \textrm{Tr}
(e^{-\beta \hat{H}})$ can be calculated making use of the $SU(1,1)$
decomposition~\cite{ban} for the exponential $\exp (-\beta \hat{H})$
in Eq.~(\ref{hamil}) and making use of the Fock basis $\vert n
\rangle$. It can be shown (see Appendix A) that this partition function may be
expressed as
\begin{equation}
Z(\beta, \hat{H})=\frac{\zeta^{1/4} e^{- \beta {\rm\, Im} (\omega_2)}}
{(1-2 \zeta^{1/2} +\zeta (1- \xi))^{1/2}} \, ,
\label{z}
\end{equation}
where the functions $\xi$ and $\zeta$ depend on the temperature, the
Hamiltonian parameters and the frequency of the quadrature
components $\omega_0$ (i.e.,  $\hat{p}=i
\sqrt{\omega_0/2}(\hat{a}^\dagger-\hat{a})$ and
$\hat{q}=(\hat{a}+\hat{a}^\dagger)/\sqrt{2\omega_0}$); these are
given by
\begin{eqnarray}
&\xi = \frac{4 \vert \gamma_1 \vert^2 \sinh^2 \phi}{\left(\omega_0
\omega_1 + \frac{\omega_3}{\omega_0}\right)^2-4 \vert \gamma_1 \vert^2} \, , \nonumber \\
&\zeta =\left(\cosh \phi +\frac{\left(\omega_0 \omega_1 + \frac{\omega_3}{\omega_0}\right)
\sinh \phi}{\left(\left(\omega_0 \omega_1 + \frac{\omega_3}{\omega_0}\right)^2-4
\vert \gamma_1 \vert^2 \right)^{1/2}}\right)^{-2} \, ,  \nonumber \\
&  \phi =  \beta \sqrt{\left(\omega_0 \omega_1 + \frac{\omega_3}{\omega_0}\right)^2-4
\vert \gamma_1 \vert^2} \, ,
\end{eqnarray}
and $\gamma_1= (\omega_3/ \omega_0 -\omega_0 \omega_1-2 i \, {\rm
Re}(\omega_{2}))/2$. It is important to notice that, in the Gaussian
case, the dimension of the Hilbert space is infinite, in such a case it
is not always possible to define the partition function of a system.
For that reason, the validity of inequality~(\ref{ineq}) is
restricted to the systems with a well-defined partition function.
For example, in the case of the harmonic oscillator, the partition
function is $Z(\beta,\omega)=\sum_{n=0}^\infty \exp (-\beta \omega
(n+1/2))$; this sum is only convergent when $T>0$, also there can be
Hamiltonians where this kind of sum is divergent even for $T>0$.

The von Neumann entropy associated to the density matrix
$\hat{\rho}$ is given by the expression~\cite{serafini}
\begin{equation}
S(\hat{\rho})=\frac{1- \mu}{2 \mu} \ln \left( \frac{1+\mu}{1-\mu}\right)
- \ln \left( \frac{2 \mu}{1+\mu} \right) \, ,
\end{equation}
where $\mu$ is the purity of the system
$\mu =(2 \sqrt{\textrm{Det}  \boldsymbol{\sigma}})^{-1}$.

\subsection{Example}
To illustrate the interpretation of Gaussian states as thermal
equilibrium states, we take a system described by the degenerated
parametric amplifier Hamiltonian~\cite{louisell}
\begin{equation}
\hat{H}= \omega_0 \left( \hat{a}^\dagger \hat{a}+\frac{1}{2}\right)
- k (\hat{a}^{\dagger 2} e^{-i \omega t}+ \hat{a}^2 e^{i \omega t}) \, ,
\label{amp}
\end{equation}
this Hamiltonian describes the interaction of a pump beam of
frequency $\omega$ with a signal beam of frequency $\omega_0$ in a
nonlinear crystal. This interaction results in the amplification of
the signal beam.

In this case, the Hamiltonian parameters of Eq.(\ref{hamil}) are
given by
\[
\omega_1=\frac{1}{2}+\frac{k}{\omega_0} \cos \omega t, \quad
\omega_2=\omega_2^*=k \sin \omega t, \quad
\omega_3=\frac{\omega_0^2}{2}-k \omega_0 \cos \omega t \, .
\]

The density matrix of the Gaussian state is given by the thermal light state
\begin{equation}
\hat{\rho}_{tl}= \sum_n \frac{\bar{n}^n}{(1+\bar{n})^{n+1}}
\vert n \rangle \langle n \vert \, ,
\end{equation}
where $\bar{n}=1/(\exp (\omega_t / T')-1)$, with $T'$ being
associated to the temperature of the light generation. It is well
known that this thermal light state can be expressed as a thermal
equilibrium state with density matrix given by $e^{-\beta
\omega_t
\hat{a}^\dagger \hat{a}}/{\rm Tr}(e^{-\beta \omega_t \hat{a}^\dagger
\hat{a}})$. For this reason, one can expect that for a small
interaction constant $k$ for the parametric amplifier Hamiltonian of
Eq.~(\ref{amp}), the parametric amplifier gives arise to a thermal
equilibrium state close to a thermal light state.

\begin{figure}
\centering
(a)
\includegraphics[scale=0.30]{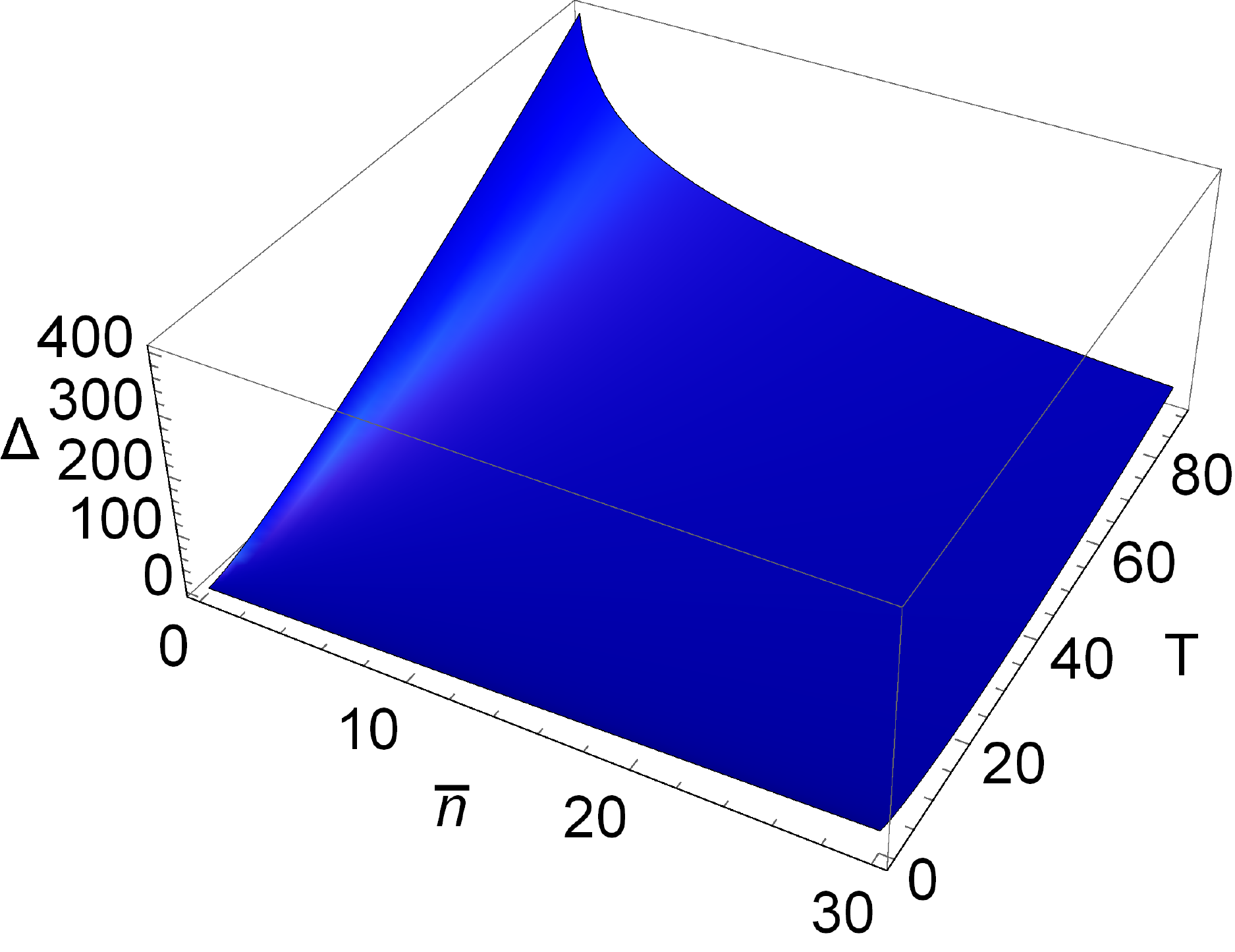} \
(b)
\includegraphics[scale=0.25]{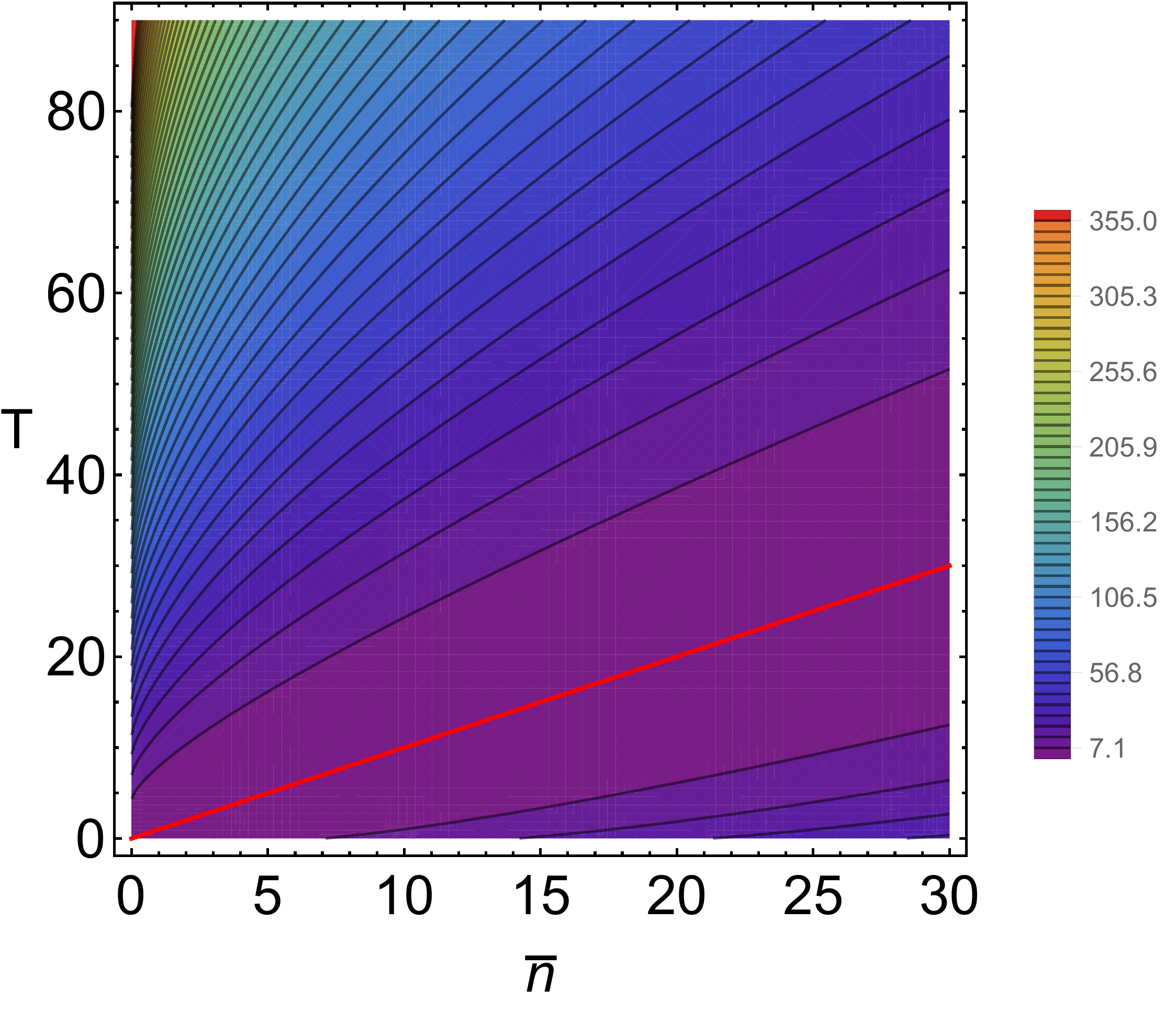}
\caption{(a) $\Delta$ parameter as a function of $T$ and $\bar{n}$
for the thermal light state and the Hamiltonian of a degenerated
parametric amplifier. (b) Contour plots of the $\Delta$ parameter,
the red line corresponds to $T=\bar{n}$. The Hamiltonian parameters
used are: $\omega_0=\omega_t=1$, $\omega=3$ and $k=0.1$.}
\label{fig:amplifier_tl}
\end{figure}

The covariance matrix for the thermal light state, the entropy and
the quadrature components mean value can be expressed as
\begin{eqnarray}
\sigma (0) =\frac{1+2 \bar{n}}{2} \left( \begin{array}{cc} \omega_0 & 0 \\ 0 & 1/\omega_0 \end{array} \right) \, , \nonumber \\
S=\bar{n} \ln \left( \frac{1+\bar{n}}{\bar{n}} \right)+\ln (\bar{n}+1)\, , \quad \langle \hat{p} \rangle= \langle \hat{q} \rangle=0 \, .
\end{eqnarray}
These expressions combined with Eqs.~(\ref{mean}) and (\ref{z}) are
used to obtain the evaluation of the parameter $\Delta$. In
figure~\ref{fig:amplifier_tl}, one can see the parameter $\Delta$ as
a function of the parameter $T$ and $\bar{n}$ for fixed Hamiltonian
with $\omega_0=1$, $\omega=3$ and $k=0.1$. As expected for small
$k$, one can see that the thermal light state can be interpreted as
a quasi-thermal equilibrium state of the parametric amplifier
Hamiltonian. We can also notice that the minimum values of $\Delta$
are located around the line $T=\bar{n}$.

\section{Summary and concluding remarks}
In this work, we analyzed the Tsallis relative entropy between the
states $\hat{\rho}$ and $\hat{\sigma}$: $S_q
(\hat{\rho},\hat{\sigma})$ for $q=1+\delta$ ($\delta \ll 1$,
$\delta>0$). The result is expressed in terms of powers of $\delta$.
Studying the first term of the series (the von Neumann relative
entropy), we defined a parameter $\Delta$ that measures the distance
between an arbitrary density $\hat{\rho}$ and a thermal equilibrium
state $\hat{\sigma}=e^{-\beta \hat{H}}/{\rm Tr}(e^{-\beta
\hat{H}})$. This parameter can be expressed in terms of the energy
$E(\hat{\rho},\hat{H})$, the entropy $S(\hat{\rho})$ and the
partition function $Z(\beta, \hat{H})$ of the system. We studied
this parameter for a general qubit system and unimodal Gaussian
state.

Also we showed that a qubit system is equal to a thermal equilibrium
state, if the Bloch vectors of the state $\mathbf{p}$ and the
Hamiltonian $\mathbf{h}$ follow the relation $\mathbf{p}=-\mathbf{h}
\tanh (\beta \vert \mathbf{h} \vert/2)/\vert
\mathbf{h} \vert$. This condition implies that the two Bloch vectors
are antiparallel for $T \geq 0$ and parallel for $T<0$. In both
cases, the temperature where $\Delta=0$ is given by $T=\vert
\mathbf{h} \vert/(2 \, { \rm arctanh} \vert \mathbf{p}\vert)$.

In the Gaussian case, the general partition function for a quadratic
Hamiltonian was obtained in order to calculate $\Delta$. As an
example, we presented the comparison between the thermal light state
and a thermal equilibrium state given by the parametric amplifier
Hamiltonian. We demonstrated that, in the case of small interaction
constant $k$ of the parametric amplifier, there is a region ($T
\approx \bar{n}$) where $\Delta$ is minimum.

\section*{Acknowledgments}
This work was partially supported by CONACyT-M\'exico (under Project
No. 238494).

\appendix
\section{Partition function for unimodal quadratic Hamiltonian systems}
The partition function associated to the arbitrary unimodal quadratic Hamiltonian of  Eq. (\ref{hamil}) is obtained in this section. It is possible to see that the quadratic Hamiltonian $\hat{H}$ can be expressed in terms of the SU(1,1) operators $\hat{K}_+=\hat{a}^{\dagger 2}/2$, $\hat{K_-}=\hat{a}^2/2$ and $\hat{K}_0=(\hat{a}^\dagger \hat{a}+1/2)/2$ as follows
\[
\hat{H}=2\left( \gamma_1 \hat{K}_- + \gamma_1^* \hat{K}_++ \left(\frac{\omega_3}{\omega_0}+\omega_0 \omega_1\right) \hat{K}_0\right)+\textrm{Im}(\omega_2),
\]
where $\gamma_1= (\omega_3/ \omega_0 -\omega_0 \omega_1-2 i \, {\rm
Re}(\omega_{2}))/2$. Because of this property the exponential operator $e^{-\beta \hat{H}}$ can be decompose as the product of exponential of each element of the algebra i.e.,
\[
e^{-\beta \hat{H}}=e^{-\beta \, \textrm{Im}(\omega_2)} e^{A_+ \hat{K}_+} e^{\ln (A_0) \hat{K}_0} e^{A_- \hat{K}_-}
\]
where 
\begin{eqnarray}
A_+ =\frac{(-2 \beta \gamma_1^*/ \phi) \sinh \phi}{\cosh \phi + (\beta(\omega_0 \omega_1+\frac{\omega_3}{\omega_0})/\phi)\sinh \phi} , \quad A_- =\frac{(-2 \beta \gamma_1/ \phi) \sinh \phi}{\cosh \phi + (\beta(\omega_0 \omega_1+\frac{\omega_3}{\omega_0})/\phi)\sinh \phi} , \nonumber \\
A_0 = \left( \cosh \phi + \frac{\beta (\omega_0 \omega_1 +\frac{\omega_3}{\omega_0})}{\phi} \sinh \phi \right)^{-2} , \quad \phi= \beta \left( \left(\omega_0 \omega_1 +\frac{\omega_3}{\omega_0}\right)^2-4 \vert \gamma_1 \vert^2 \right)^{1/2} \nonumber .
\end{eqnarray}
The diagonal matrix elements in the Fock representation of the exponential operator can be expressed in terms of the Legendre polynomials $P_n (z)$
\[
\langle n \vert e^{-\beta \hat{H}} \vert n \rangle=e^{-\beta \, \textrm{Im}(\omega_2)} A_0^{1/4}(A_0(1-\xi))^{n/2} P_n \left( \frac{1}{\sqrt{1-\xi}} \right), \quad \xi=\frac{A_+ A_-}{A_0} , 
\]
summing this elements over $n$ we finally have
\[
Z(\beta,\hat{H})= \frac{A_0^{1/4} e^{-\beta \, \textrm{Im}(\omega_2)}}{\sqrt{1-2 A_0^{1/2}+A_0 (1-\xi)}},
\]
renaming $A_0=\zeta$ we arrive to the expression for the partition function of Eq. (\ref{z}).

\section*{References}

\bibliographystyle{elsarticle-num}
\bibliography{mybib}

\end{document}